\documentclass[%
 aps,
 amsmath,amssymb,
 aps,
]{revtex4}
\usepackage{amssymb}
\usepackage{amsfonts}
\usepackage{amsmath}
\usepackage{epsfig, color}

\usepackage{graphicx}
\graphicspath{%
    {converted_graphics/}
    {/}
}
\usepackage{comment}
\usepackage{amsfonts,amssymb,stmaryrd,amsthm}
\usepackage{amsmath}
\usepackage{graphicx}
\usepackage{enumerate}
\usepackage{subfigure}
\usepackage{amscd}
\usepackage{pb-diagram}
\usepackage[toc]{appendix}
\usepackage{afterpage}
\usepackage{bm}
\usepackage{epstopdf}
\usepackage{mathrsfs}
\newcommand{\eproof}{\hfill{\vrule height5pt width5pt depth0pt}\medskip}

\renewcommand{\proof}{{\noindent {\em Proof}.\;}}

\newtheorem{theorem}{Theorem}[section]

\newtheorem{definition}[theorem]{Definition}

\begin{document}
\title{Stability of solitary and cnoidal traveling wave  solutions for a fifth order  Korteweg-de Vries  equation}


 \author{Ronald Adams}
  \email{adamsr25@erau.edu}

\author{Stefan C. Mancas}
\email[Electronic address for correspondence: ]{mancass@erau.edu}

\affiliation{Department of Mathematics, Embry-Riddle Aeronautical University,\\ Daytona Beach, FL 32114-3900, USA}


\begin{abstract}
We establish the nonlinear stability of solitary waves  (solitons) and periodic traveling wave solutions (cnoidal waves) for  a  Korteweg-de Vries (KdV) equation which includes a fifth order dispersive  term. The traveling wave solutions which yield solitons for zero boundary conditions and  wave-trains of  cnoidal waves for nonzero boundary conditions are analyzed using   stability theorems, which   rely on the positivity properties of the Fourier transforms. We show that all families of solutions considered here are (orbitally) stable. 

\medskip{}

\end{abstract}
\maketitle

\section{Introduction}

We investigate the stability of the traveling wave solutions    Korteweg-de Vries (KdV) equation with a fifth order dispersive term \cite{Man3,Here}. Numerically it was investigated in a study of magneto-acoustic waves in a cold collision-free plasma \cite{Kawa5}, and  takes the form 
\begin{equation}\label{eq:rosenau}
u_{t}+\gamma uu_{x}+\alpha u_{xxx}=\beta u_{xxxxx}.
\end{equation}
$\alpha, \beta$ are the third and fifth order dispersive terms, respectively, $\gamma$  is a  wave steepening parameter  for a unidirectional traveling wave  $u(\xi)=u(x-ct)$,  of  velocity $c$ in the $x$ direction at time $t$ which satisfies
 the first  conservation law
\begin{equation}\label{eq1aa}
-cu+\frac \gamma 2 u^2 +\alpha u_{\xi \xi} -\beta u_{\xi \xi \xi \xi}  =\mathcal{A},
\end{equation}
and by  multiplying by $u_\xi$, integrating and using \eqref{eq1aa} again, leads to  the second conservation law 
\begin{equation}\label{eq1ab}
-\frac c 2 u^2+\frac \gamma 3  u^3 +\alpha \left(u u_{\xi \xi}-\frac 1 2 {u_\xi}^2\right) -\beta \left(uu_{\xi \xi \xi \xi}-u_\xi u_{\xi \xi \xi}+\frac 1 2 {u_{\xi\xi}}^2\right)  =\mathcal{B}.
\end{equation}
The constants  $\mathcal{A},\mathcal{B}$ can be regarded as the mass and energy flux, and should be both zero for solitary  waves and nonzero for oscillatory tails \cite{Chow}.

We  establish the orbital stability for traveling wave solutions of  (\ref{eq:rosenau}), which depend on the parameter $\gamma$, 
by applying an existing stability criterion found in \cite{Albert} and \cite{PN}.  
Previously, the stability for solitary wave  solutions   to the initial value problem for the KdV equation was already established by Benjamin \cite{TB}, but it did not include the fifth order dispersive  term, though the literature is rich  of papers  \cite{Albert3,BBB,BSS,MW2}  that develop sufficient conditions which imply stability for long wave equations with a general linear dispersion term $u_t+u_x+u^pu+\mathcal{L}u_x=0$. The case where $\gamma=\alpha=1$ is addressed in \cite{Natali}, and the case where $\beta=\frac{1}{1680}$, $\gamma=1$, and $\alpha=\frac{13}{420}$ is studied in \cite{Albert}. 
In \cite{Albert3, BBB}, the computation of the spectrum of $\mathcal{L}$ and the ability to verify an inequality involving the eigenfunctions for $\mathcal{L}$ is required, whereas in \cite{Albert} the positivity of the Fourier transform of the solitary wave is used in conjunction with the inner product  $I=\left(\chi,\varphi\right)_{L^2(\mathbb{R})}$ to obtain stability.  Here, the class PF(2) \cite{Albert2,Karlin} is used to determine the necessary spectral properties of $\mathcal{L}$.  For the theory of instability of solitary waves we refer to the papers \cite{BD,PN3}, where the authors establish instability for solitary waves associated to a generalized fifth order KdV equation of the form $u_t+u_{xxxxx}+bu_{xxx}=(G(u,u_x,u_{xx}))_x$ for $b\neq0$, where $G(q,r,s)=F_q(q,r)-rF_{qr}(q,r)-sF_{rr}(q,r)$ and  $F(q,r)$ a homogeneous function of degree $p+1$ for $p>1$. When considering periodic traveling waves (of cnoidal type) we note that the literature is not as extensive as is the case for solitary waves.  Pomeau et al. \cite{Pom} compute the amplitudes of continuous-wave tails in the fifth-order Korteweg-de Vries equation in order to discuss the structural stability of the KdV equation under singular perturbation, while  Pava and Natali \cite{PN} provide a program well suited for addressing the issue of stability for the periodic waves considered herein. There, the explicit expression for the periodic wave is based on the Jacobi elliptic functions and the Fourier series representation thereof.  The stability theory is established under the conditions that $\widehat{\varphi_c}>0$, $\widehat{\varphi^p_c}\in PF(2)$, and $I=\left(\chi,\varphi\right)_{L^2_{2L}}<0$. The first case of proving stability of periodic traveling waves of  the KdV equation appears in \cite{MK} where the integrability of the KdV equation is exploited.  In \cite{PN2} the authors adapt the modern theory of stability of solitary waves \cite{JLB,MW2} to the periodic context.  They show that  periodic waves corresponding to the KdV equation are stable with respect to small, periodic perturbations in the context of the initial-value problem.  In \cite{GS} the authors consider an abstract Hamiltonian system in the presence of symmetry and which, relying on a sharp identification
of the lowest eigenvalues in the spectrum of the linearized problem, reduces the stability issue to verifying
the convexity of a specific functional dependent on the wave speed.

\section{Stability of analytical solutions}
In this section we discuss the traveling wave solutions along with their stability properties. 
 First,  we consider the case where the traveling wave is a solitary wave,  whereas the case of periodic traveling waves are dealt with in the subsequent subsection.
\subsection{Solitary waves}\label{subsect_1}
Assuming $\alpha$, $\beta>0$,  and using zero boundary conditions, the solution to 
$(\ref{eq:rosenau})$  which was  obtained previously  by  Hereman et al. \cite{Here} using the direct algebraic method, and more recently   by Mancas \cite{Mancass} using the elliptic function method, takes the form
\begin{equation}\label{lab9}
u(x,t)= \frac{105 \alpha^2}{169 \gamma \beta}\mathrm{sech}^4\left[\frac 1 2 \sqrt{\frac{\alpha}{13 \beta}}\left(x-\frac{36 \alpha^2}{169\beta} t\right)\right]\equiv \varphi_{c}(\xi)
\end{equation}
and represents a  solitary wave that  translates to the right with velocity   $c=\frac{36\alpha^2}{169 \beta}$   that is fixed by both dispersion coefficients. 

For the special case of  $\beta=0$,  (\ref{eq:rosenau})  reduces to the well-known  KdV equation which   describes the motion of  small amplitude and large wavelength shallow waves in dispersive systems \cite{KdV} 
\begin{equation}\label{eq:KdV}
u_{t}+\gamma uu_{x}+\alpha u_{xxx}=0,
\end{equation}  
with mass and energy flux  given by
\begin{equation}\label{eq1aaa}
\begin{array}{l}
-cu+\frac \gamma 2 u^2 +\alpha u_{\xi \xi}=\mathcal{A},\\
-\frac c 2 u^2+\frac \gamma 3  u^3 +\alpha \left(u u_{\xi \xi}-\frac 1 2 {u_\xi}^2\right) =\mathcal{B}.
\end{array}
\end{equation}
It is worth noting that for \eqref{eq:KdV} a more general case can be adopted,   that is by inclusion of the linear term $C u_x$
\begin{equation}\label{eq:gKdV}
u_{t}+C u_x+\gamma uu_{x}+\alpha u_{xxx}=0
\end{equation}  
which  appears in the work of \cite{Craig2}, and more recently in \cite{Const1,Const3}.  In \cite{Const1} a Hamiltonian formulation is given for the governing equations describing the two-dimensional nonlinear interaction between coupled surface waves, internal waves, and an underlying current with piecewise constant
vorticity in a two-layered fluid overlying a flat bed. In \cite{Craig2} the authors develop a Hamiltonian perturbation theory for the long-wave limits, and carry out analysis of the principal long-wave scaling regimes for irrotational flows (with zero vorticity). Note that in the presence of non-uniform underlying currents there is non-zero vorticity, and the most wide ocean motion with coherent travelling waves occurs in the equatorial Pacific (over more than $12,000$ km.), where underlying currents are of great significance \cite{Const3}. The revised model (\ref{eq:gKdV}) can be reduced to the original Eq.  \eqref{eq:KdV} by looking at a frame moving at a suitable constant speed,  that is, after a change of variables $(x,t)$ to $(x-C t,t)$.

The solution of  \eqref{eq:KdV} using zero boundary conditions is the solitary wave \cite{KdV,Rey,Mancass}
\begin{equation}\label{kdv3}
u(x,t) =\frac{3c}{\gamma}\mathrm{sech}^2\left[\frac 1 2\sqrt{\frac{c}{\alpha}}(x-ct)\right]\equiv  \phi_c(\xi)
\end{equation}
which propagates with arbitrary velocity  to the right  $c>0$ when $\alpha>0$ or to the left $c<0$ when $\alpha<0$.

We first establish the stability of solitary waves  for  (\ref{eq:KdV}) using $(\ref{kdv3})$ and then for  (\ref{eq:rosenau}) using (\ref{lab9}), with suitable conditions on the parameters $\gamma$, $\alpha$, $\beta$.  Throughout we will make use of the results and definitions found in \cite{Natali,Albert}.  In  \cite{Albert} the author remarks that the results therein can be extended to include  more general nonlinearities, and with this in mind we consider a version of Eq. (1.1) in \cite{Albert} with an appended factor of $\gamma$ to the nonlinear term $u^{p}u_{x}$ that takes the form
\begin{align}\label{eq:pde2}
u_{t}+\gamma uu_{x}-\left(Mu\right)_{x}=0,
\end{align}
where $M$ is a differential  operator with positive symbol defined by $M=\beta \frac{d^{4}}{dx^4}-\alpha\frac{d^3}{dx^3}$.

Next, we establish the stability of the  family of solutions given by (\ref{kdv3}) by applying Theorem~3.1 in \cite{Albert}.  Using the traveling wave  ansatz  and integrating once  (\ref{eq:pde2}) assuming zero integration constant, we obtain
\begin{align}\label{eq:ode1}
(M+c)u-\frac{\gamma}{p+1}u^{p+1}=0.
\end{align}
We define a solitary wave $\varphi$ as an even function which lies in the space $H^{\mu/2}$ and is a solution to  (\ref{eq:ode1}).  To study the stability of traveling waves for (\ref{eq:rosenau}) we must consider the associated linear operator
$\mathcal{L}:L^2(\mathbb{R})\rightarrow L^2(\mathbb{R})$
\begin{align}\label{eq:L_operator}
\mathcal{L}\zeta=\left(M+c\right)\zeta-\gamma u^p\zeta.
\end{align}
Proposition~2.1 in \cite{Albert} establishes that $\mathcal{L}$ is a linear, closed, unbounded, self-adjoint operator defined on a dense subspace of $L^2(\mathbb{R})$.  In particular this means $\mathcal{L}$ has the eigenvalue 0, with eigenfunction $\frac{du}{dx}$.  
\begin{definition}\label{deff}
\em{ Let $\varphi$ be a solitary traveling-wave solution of (\ref{eq:rosenau}) and consider $\tau_{r}\varphi(x)=\varphi(x+r)$, $x,r\in\mathbb{R}$.  We define the set $\Omega_{\varphi}\subset H^2(\mathbb{R})$ the orbit generated by $\varphi$, as 
\begin{align*}
\Omega_{\varphi}=\left\{g\mid g=\tau_r\varphi, \text{for some $r\in\mathbb{R}$}\right\}.
\end{align*}
Moreover, for any $\eta>0$, define $U_{\eta}\subset H^2(\mathbb{R})$ by 
$$U_{\eta}=\left\{f:\inf_{g\in\Omega_{\varphi}}\left\|f-g\right\|_{H^2}<\eta\right\}.$$
Using this terminology $\varphi$ is said to be \textit{(orbitally) stable} if
\begin{enumerate}\label{def:stability}
\item[(i)] the initial value problem associated with  (\ref{eq:rosenau}) is globally well-posed in $H^2(\mathbb{R})$.
\item[(ii)] For every $\epsilon>0$, there exists $\delta>0$ such that for all $u_0\in U_{\delta}$, the solution $u$ to  (\ref{eq:rosenau}) with initial condition  $u(0,x)=u_0$ satisfies $u(t)\in U_{\epsilon}$ for all $t>0$.
\end{enumerate} 
}\end{definition}  
\begin{theorem}\label{Thm1}\em{
The smooth family of solutions (\ref{kdv3}) is orbitally stable in $H^2(\mathbb{R})$ under the flow of  \eqref{eq:KdV}.
}\end{theorem}
\proof Recall the linear operator associated to (\ref{eq:KdV})
\begin{align}
\mathcal{L}=\left(-\alpha \frac{d^2}{dx^2}+c\right)-\gamma\phi_{c},
\end{align}
with $M=M_{1,1}=-\alpha \frac{d^2}{dx^2}$, therefore by Theorem~4.6 in \cite{Albert} it suffices to show that 
 $I=\left(\phi_c,\psi\right)_2<0$.  To do this we compute $\frac{d}{dc}\left\|\phi_c\right\|^2$, where 
\begin{align}
\nonumber\left\|\phi_c\right\|^2&=\left(\frac{3c}{\gamma}\right)^2\int_{\mathbb{R}}\mathrm{sech}^4 \left(\frac{1}{2} \sqrt{\frac{c}{\alpha}}\xi\right)\;d\xi=\left(\frac{3c}{\gamma}\right)^22\sqrt{\frac{\alpha}{c}}\int_{\mathbb{R}}\mathrm{sech}^4 (\chi)\;d\chi=\left(\frac{3c}{\gamma}\right)^22\sqrt{\frac{\alpha}{c}}\left(\frac{4}{3}\right)\\&=\frac{24\alpha^{1/2}}{\gamma^2}c^{3/2}.
\end{align}
Therefore $\frac{d}{dc}\left\|\phi_c\right\|^2>0$, for all $c>0$.\eproof

Next, we establish the stability of the traveling wave solution (\ref{lab9}).   Since (\ref{lab9}) does not define a family of solutions in $c$, we make use of Gegenbauer polynomials \cite{Albert}.  We can use these polynomials to help determine the sign of the inner product   $I=\left(\varphi,\psi\right)_2$.  Specifically, we use Theorem 4.10 in \cite{Albert} which provides us with the proper expression for $I$ in terms of the gamma function which takes the form
\begin{align}
I=a\sum^{\infty}_{j=0}\left(\frac{\lambda_{2j}}{1-\lambda_{2j}}\right)\left\{\frac{\Gamma(2j+1)\cdot(2j+n+r-\frac{1}{2})}{\Gamma(2j+2n+2r-1)}\right\}\left\{\frac{\Gamma(j+n)\Gamma(j+n+r-\frac{1}{2})}{\Gamma(j+1)\Gamma(j+r+\frac{1}{2})}\right\}^2,
\end{align}
where $a=\left(\frac{\gamma 2^{n+r-1}\Gamma(r)}{\pi\Gamma(n)}\right)$, $\lambda_m=\frac{\Gamma(r+m)}{\Gamma(r+1)}\cdot \frac{\Gamma(r+2n+1)}{\Gamma(r+2n+m)}$, $r=4$ and $n=2$.  Let $b_j$ represent the $j$th term of the series, since $b_0<0$ it suffices to show that $\sum^{\infty}_{j=1}b_j<\left|b_0\right|$, where 
\begin{align*}
b_j=\frac{1680\left(2j+\frac{11}{2}\right)\left(j+1\right)^2\left(j+\frac{9}{2}\right)^2\left(2j\right)!}{\left[(2j+4)(2j+5)(2j+6)(2j+7)-1680\right](2j+10)!},
\end{align*}  and 
$\left|b_0\right|=\left(\frac{11}{10!}\right)\left(\frac{81}{4}\right)\approx 6.14\times 10^{-5}$. By the use of Stirling's formula $b_j\sim j^{-2r-1}$ as $j\rightarrow\infty$, hence we have $\sum^{\infty}_{j=1}b_j\approx5.05\times10^{-6}<\left|b_0\right|$.
\begin{theorem}{\em
The smooth family of solutions (\ref{lab9}) is orbitally stable in $H^2(\mathbb{R})$ under the flow of \eqref{eq:rosenau}.}
\end{theorem}
First we note that the differential operator corresponding to \eqref{eq:rosenau} is given by $M_{2,1}=\beta \frac{d^4}{dx^4}-\alpha \frac{d^2}{dx^2}$.  We again appeal to Theorem~4.6 in \cite{Albert}, hence it remains to show that $I=(\varphi,\psi)<0$ which was already established above.  \eproof

\subsection{Periodic traveling waves}\label{per}
For the KdV Eq.   \eqref{eq:KdV} with nonzero boundary conditions the solution found in  \cite{Mancass}  is 
\begin{equation}\label{Tab1}
  u(x,t) =  \begin{array}{lr}
\frac{3c + \sqrt{\Delta}}{2\gamma} \mathrm{cn}^2\left[\frac 1 2 \frac{\sqrt[4]{\Delta}}{\sqrt{3\alpha}}(x-ct);\sqrt{\frac 1 2 \left(1+ \frac{3c}{\sqrt{\Delta}}\right)}\right]\equiv  \varphi_c(\xi),
 \end{array}
\end{equation} which can be written compactly as
\begin{equation}\label{so}
\varphi_c(\xi)=A~\mathrm{cn}^2\left(\frac 1 2 \frac{\sqrt[4]{\Delta}}{\sqrt{3\alpha}}\xi;k\right),
\end{equation}
 where $\mathrm{cn}(\theta;k)$ is the Jacobian elliptic function
with amplitude $A=\frac{3c + \sqrt{\Delta}}{2\gamma}$, and modulus $k=\frac{\sqrt{A \gamma}}{\sqrt[4]\Delta}$. This solution  represents a train of periodic cnoidal waves which propagate with arbitrary velocity  and  wavelength  given by  
\begin{align}\label{wl}
\lambda=\frac{4\sqrt{3\alpha}}{\sqrt[4]{\Delta}}K(k),
\end{align}
where  $\Delta=9c^2+24 \mathcal {A} \gamma>0$,  and $K(k)$ is the complete elliptic integral of the first kind \cite{Steg}  $K(k)= \int_0^{\frac{\pi}{2}}\frac{d\theta}{\sqrt{1-k^2 \sin^2\theta}}.$

For the special case  $\alpha=0$, $\beta \ne 0$,  (\ref{eq:rosenau})  takes the form
\begin{equation}\label{eq:naga}
u_{t}+\gamma uu_{x}=\beta u_{xxxxx}.
\end{equation}
This equation  was first studied by Hasimoto \cite{Has} for shallow water waves near some critical value of surface tension, while  Nagashima \cite{Naga1} performed experiments, and observed  solitary waves with small oscillating tails using an oscilloscope. Its conservation laws  are
\begin{equation}\label{eq20aa}
\begin{array}{l}
-cu+\frac \gamma 2 u^2 -\beta u_{\xi \xi \xi \xi}  =\mathcal{A}, \\
-\frac c 2 u^2+\frac \gamma 3  u^3  -\beta \left(uu_{\xi \xi \xi \xi}-u_\xi u_{\xi \xi \xi}+\frac 1 2 {u_{\xi\xi}}^2\right)  =\mathcal{B}.
\end{array}
\end{equation}
The solution to  (\ref{eq:naga}) obtained also by \cite{Mancass,Yama7} is
\begin{equation}\label{eq005}
u(x,t)= \frac{5c}{2\gamma} ~\mathrm{cn}^4\left[\frac{\sqrt 2}{2}\sqrt[4]{\frac{c}{42 \beta}}~(x-ct);\frac{\sqrt 2}{2}\right] \equiv \phi_c(\xi),
\end{equation}
 and  represents a train of periodic cnoidal waves which only propagate to the right with   shape  preserved by the constant modulus, and  wavelength  given by $
 \lambda=2 \sqrt 2 \sqrt[4]{\frac{42 \beta}{c}} K\left(\frac {\sqrt 2}{ 2}\right).$
 
Next, we establish the stability of the periodic traveling-wave solutions given by (\ref{Tab1}) and  (\ref{eq005}). For this we make use of the techniques developed in \cite{PN}.  This requires the following adjustments to our current setup.  That is, we consider traveling-wave solutions to  (\ref{eq:rosenau}) of the form $u(x,t)=\varphi_{c}(x-ct)$, where the profile $\varphi_{c}$ is a smooth periodic function with fundamental period $\lambda=2L$ given by $(\ref{wl})$  for $L>0$. 
 
The notion of stability carries over for periodic traveling waves in this context,  that is we say the orbit generated by $\varphi_c$ denoted $\Omega_{\varphi_c}=\left\{\varphi_c(\cdot+y)\mid y\in\mathbb{R}\right\}$ is stable in $H^{m_2}_{per}\left([-L,L]\right)$ by the periodic flow generated by  $(\ref{eq:rosenau})$.  If, for any $\epsilon>0$, there exists a $\delta>0$ such that for $u_0\in H^{m_2}_{per}\left([-L,L]\right)$ with $d(u_0,\Omega_{\varphi_c})\equiv \inf_{y\in\mathbb{R}}\left\|u_0-\varphi_c(\cdot+y)\right\|_{H^{m_2}_{per}}<\delta$.  The solution $u$ of  (\ref{eq:rosenau})  with $u(x,0)=u_0$ is global in time and satisfies $d\left(u(\cdot,t),\Omega_{\varphi_c}\right)<\epsilon$ for all $t\in\mathbb{R}$.  With this notion of stability the conditions found in the papers \cite{PN,TB,JLB,MW,GS} can be used to imply stability:
\begin{enumerate}
\item[$(P_0)$] There is a nontrivial smooth curve of periodic solutions for  $(\ref{eq:rosenau})$ of the form $c\in I\subset\mathbb{R}\rightarrow\varphi_c\in H^{m_2}_{per}\left([-L,L]\right)$,
\item[$(P_1)$] $\mathcal{L}$ has a unique negative eigenvalue $\lambda$ and it is simple,
\item[$(P_2)$] the eigenvalue is 0,
\item[$(P_3)$] $\frac{d}{dc}\int^{L}_{-L}\varphi^2_c(x)dx>0.$
\end{enumerate}
By considering the periodic solutions (\ref{Tab1}) and (\ref{eq005}) condition $(P_0)$ will be satisfied.  To check conditions $(P_1)$ and $(P_2)$ we use Theorem~4.1 in \cite{PN} which relies on the positivity properties of the Fourier transform of the solution.
 The main theorems used to verify conditions $(P_0)-(P_2)$ are Theorem~5.1 and Theorem~4.1 in \cite{PN}.

The first periodic traveling wave solution we consider is  (\ref{Tab1}) which is a solution to  \eqref{eq:KdV}
with condition $\mathcal{A}\gamma>0$.  Using the wavelength formula given by (\ref{wl}) with  $L=\frac \lambda 2$, we can write   $(\ref{Tab1})$ in a simpler form
\begin{align}\label{per_trav2}
\varphi_c(\xi)= \frac{2\mathcal{M}(c)K^2(k)}{L^2}\mathrm{cn}^2\left[\frac{K(k)}{L}\xi;k\right],
\end{align}
where $\mathcal{M}(c)=\frac{3\alpha}{\gamma}\left(1+\frac{3c}{\sqrt{\Delta}}\right)=\frac{6 \alpha A}{\sqrt{\Delta}}>0.$
\begin{theorem}
The periodic traveling wave solution $(\ref{per_trav2})$ is stable in $H^1_{per}([0,L])$ by the flow of Eq. $(\ref{eq:naga})$.
\end{theorem}
\proof  We first consider the Fourier expansion of $\text{cn}^2(\cdot,k)$
\begin{align}\label{Fourier_exp}
\text{cn}^2\left(\frac{K}{L}\xi;k\right)=1-\frac{1}{k^2}\left(1-\frac{E}{K}\right)+\frac{2\pi^2}{k^2K^2}\sum^{\infty}_{n=1}\frac{nq^n}{1-q^{2n}}\cos\left(\frac{n\pi}{L}\xi\right),
\end{align}
where $E$ is the complete integral of the second kind \cite{Steg} $$E(k)=\int^{\pi/2}_0\sqrt{1-k^2\sin^2(\theta)}\;d\theta.$$  The series in $(\ref{Fourier_exp})$ converges when the nome $q=e^{-\frac{\pi K}{K^{\prime}}}$ satisfies $qe^{2\text{Im}(\zeta)}<1$ where $\zeta=\frac{\pi \xi}{L}$.  Since $\text{Im}(\zeta)=0$ for $\xi\in\mathbb{R}$ and $q<1$ we see that the series $(\ref{Fourier_exp})$ converges.  Furthermore  
\begin{align}
\frac{q^n}{1-q^{2n}}=\frac{1}{2}~\mathrm{csch}\left(\frac{n\pi K^{\prime}}{K}\right),
\end{align}
where $K^{\prime}(k)=K(k^{\prime})$ and $k^{\prime}=\sqrt{1-k^2}$.  Therefore
\begin{align}
K^2\text{cn}^2\left(\frac{K}{L}\xi;k\right)=K^2-K\frac{\left(K-E\right)}{k^2}+\frac{\pi^2}{k^2}\sum^{\infty}_{n=1}n~\mathrm{csch}\left(\frac{n\pi K^{\prime}}{K}\right)\cos\left(\frac{n\pi}{L}\xi\right).
\end{align}
From this we obtain that the Fourier coefficients of $\varphi_c$ are
\begin{align}\widehat{\varphi_c}(n)=
\begin{cases}
\frac{2\mathcal{M}K}{L^2}\left(K-\frac{K-E}{k^2}\right),\quad n=0\\
\frac{2\mathcal{M}\pi^2}{L^2k^2}n~\mathrm{csch}\left(\frac{n\pi K^{\prime}}{K}\right),\quad n\neq 0.
\end{cases}
\end{align}
The expression $\frac{2\mathcal{M}K}{L^2}\left(K-\frac{K-E}{k^2}\right)$ is positive on $(0,1)$ (this is discussed below) therefore $\varphi_c>0$.  Moreover, we see that $\widehat{\varphi}_c>0$ due to the Fourier coefficients of $\varphi_c$.  

We consider the function $f:\mathbb{R}\rightarrow\mathbb{R}$ defined by $f(x)=\frac{2\mathcal{M}\pi^2}{L^2k^2}x~\mathrm{csch}\left(\frac{\pi x K^{\prime}}{K}\right)$. To show that $\widehat{\varphi_c}$ belongs to $PF(2)$ in the discrete case we define the function $h:\mathbb{R}\rightarrow\mathbb{R}$ with  $h(0)=\frac{2\mathcal{M}K}{L^2}\left(K-\frac{K-E}{k^2}\right)$, and $h(x)=f(x)$ for $x\in\left(-\infty,-1\right]\cup\left[1,\infty\right)$.  On $(-1,1)$ we extend $f$ in a differentiable manner such that $h(x)$ belongs to $PF(2)$ continuous case.  Hence, in the discrete case $h(n)=\widehat{\varphi_c}(n)$ is in $PF(2)$ discrete.  By Theorem~4.1 in \cite{PN} properties $(P_1)$ and $(P_2)$ in Definition~5.1 are satisfied. 
Next, we set $\chi=-\frac{d}{dc}\varphi_c$ since $\mathcal{L}\chi=\varphi_c$, by Parseval's theorem it follows that 
\begin{align*}
I=-\frac{L}{2}\frac{d}{dc}\left\|\varphi_c\right\|^2_{L^2_{per}}=-\frac{L}{2}\frac{d}{dc}\left\|\widehat{\varphi_c}\right\|^2_{\ell^2},
\end{align*}
where
\begin{align*}
\left\|\widehat{\varphi_c}\right\|^2_{\ell^2}=\frac{4\mathcal{M}^2K^2}{L^4}\left(K-\frac{K-E}{k^2}\right)^2+\frac{4\mathcal{M}^2\pi^4}{L^4k^4}\sum_{n\neq0}n^2~\mathrm{csch}^2\left(\frac{n\pi K^{\prime}}{K}\right),
\end{align*}
and $D(k)=\frac{K(k)-E(k)}{k^2}$ is the Legendre integral given by $D(k)=\int^{\pi/2}_{0}\frac{\sin^2(\theta)}{\sqrt{1-k^2\sin^2(\theta)}}d\theta$.

Therefore, 
\begin{align}\nonumber
&\frac{d}{dc}\left\|\widehat{\varphi_c}\right\|^2_{\ell^2}=\frac{4\mathcal{M}K}{L^4}\left(K-D\right)^2\frac{d}{dc}(\mathcal{M}K)+\frac{8\mathcal{M}^2K^2}{L^4}\left(K-D\right)\frac{d}{dc}\left(K-D\right)\\&\nonumber+\frac{4\pi^4}{L^4}\left(\frac{k\frac{d\mathcal{M}}{dc}-2\mathcal{M}\frac{dk}{dc}}{k^3}\right)\sum_{n\neq0}n^2~\mathrm{csch}^2\left(\frac{n\pi K^{\prime}}{K}\right)\\&+\frac{8\pi^5}{L^4}\left(\frac{\mathcal{M}}{k^2}\right)^2\left(\frac{K^{\prime}\frac{dK}{dk}-K\frac{dK^{\prime}}{dk}}{K^{2}}\right)\frac{dk}{dc}\sum_{n\neq0}n^3~\mathrm{csch}^2\left(\frac{n\pi K^{\prime}}{K}\right)\coth\left(\frac{n\pi K^{\prime}}{K}\right).
\end{align}
To determine the sign of $\frac{d}{dc}\left\|\widehat{\varphi_c}\right\|^2_{\ell^2}$ we consider the following terms:

\begin{enumerate}
\item [(i)] $\frac{4\mathcal{M}K}{L^4}\left(K-D\right)^2\frac{d}{dc}(\mathcal{M}K),$
\item [(ii)]$\frac{8\mathcal{M}^2K^2}{L^4}\left(K-D\right)\frac{d}{dc}\left(K-D\right),$
\item[(iii)]$\frac{4\pi^4}{L^4k^3}\left(k\frac{d\mathcal{M}}{dc}-2\mathcal{M}\frac{dk}{dc}\right),$
\item[(iv)]$\frac{8\pi^5}{L^4}\left(\frac{\mathcal{M}}{k^2}\right)^2\left(\frac{K^{\prime}\frac{dK}{dk}-K\frac{dK^{\prime}}{dk}}{K^{2}}\right)\frac{dk}{dc}.$
\end{enumerate}
\noindent
For (i), note that $\mathcal{M}(c)$, $K(c)\geq 0$, furthermore $\mathcal{M}(c)^{\prime}=\frac{216\left|\alpha\right|\mathcal{A}\gamma}{\left[9c^2+24\mathcal{A}\gamma\right]^{3/2}}\geq0$, $\frac{dK}{dc}=\frac{dK}{dk}\frac{dk}{dc}>0$. \\
 For (ii),  note that $K(k)-D(k)=\int^{\pi/2}_0\frac{1-\sin(\theta)}{\sqrt{1-k^2\sin^2(\theta)}}d\theta>0$ and also $\frac{d}{dc}\left(K-D\right)>0$. \\
 For (iii),  note that $\mathcal{M}=\frac{6\left|\alpha\right|k^2}{\gamma}$ 
 hence $k\frac{d\mathcal{M}}{dc}-2\mathcal{M}\frac{dk}{dc}=\frac{3\left|\alpha\right|}{\gamma}\left(4k^2\frac{dk}{dc}-4k^2\frac{dk}{dc}\right)=0.$ \\
 Lastly, for (iv), since $K^{\prime}$, $\frac{dK}{dk}>0$ and $\frac{dK^{\prime}}{dk}<0$ we have $K^{\prime}\frac{dK}{dk}-K\frac{dK^{\prime}}{dk}>0$.  Therefore, $I=-\frac{d}{dc}\left\|\widehat{\varphi_c}\right\|^2_{\ell^2}<0$ hence,  by Theorem~5.1 in \cite{PN} the positive cnoidal waves $\varphi_{c}$ are stable in $H^1_{per}\left(\left[0,L\right]\right)$. \eproof

Finally,  we turn to the question of stability for the periodic traveling wave solution (\ref{eq005}) of  \eqref{eq:naga}.
\begin{theorem}
The periodic traveling wave solution $(\ref{eq005})$ is stable in $H^1_{per}([0,L])$ by the flow of Eq.  $(\ref{eq:naga})$.
\end{theorem}
\proof
In order to proceed, we first consider the Fourier series expansion for $\mathrm{cn}^4$, see \cite{AK}.   
\begin{align}\label{cn4_exp}
k^4\mathrm{cn}^4(z,k)=\frac{1}{3}\left[2\left(k^2-k^{\prime2}\right)\left(\frac{E}{K}-k^{\prime2}\right)+k^2k^{\prime2}\right]+\frac{2\pi^2}{K^2}\sum^{\infty}_{n=1}\frac{nq^n}{1-q^{2n}}\frac{1}{3}\left(2(k^2-k^{\prime2})+\frac{n^2\pi^2}{2K^2}\right)\cos\left(\frac{n\pi}{K}z\right),
\end{align}
where $z=\frac{\sqrt 2}{2}\sqrt[4]{\frac{c}{42 \beta}}~\xi$, the series converges since  $0=\text{Im}(z/K)<\text{Im}(iK^{\prime}/K)$. Moreover, for $k=\frac{\sqrt{2}}{2}$   (\ref{cn4_exp})  reduces to 
\begin{align}
\mathrm{cn}^4(z,\sqrt{2}/2)=\frac{1}{3}+\frac{2\pi^4}{3K^4}\sum^{\infty}_{n=1}n^3~\mathrm{csch}(n\pi)\cos\left(\frac{n\pi}{K}z\right).
\end{align}
From this we compute the Fourier coefficients of $\phi_c$  which  are
\begin{align}\widehat{\phi_c}(n)=
\begin{cases}
\frac{5c}{6\gamma},\quad n=0\\
\frac{5c\pi^4}{3\gamma K^4}n^3~\mathrm{csch}(n\pi),\quad n\neq 0.
\end{cases}
\end{align}
Since $\widehat{\phi_c}(n)>0$, following the same argument as in the previous proof we can conclude that $\widehat{\phi_c}(n)$ is in $PF(2)$ discrete. 
Let $\chi=-\frac{d}{dc}\phi_c$ then $\mathcal{L}\chi=\phi_c$, hence, by Parseval's theorem it follows that $I=-\frac{L}{2}\frac{d}{dc}\left\|\phi_c\right\|^2_{L^2_{per}}=-\frac{L}{2}\frac{d}{dc}\left\|\widehat{\phi_c}\right\|^2_{\ell^2}$, thus
\begin{align*}
\left\|\widehat{\phi_c}\right\|^2_{\ell^2}=\frac{25c^2}{36\gamma^2}+\frac{25c^2\pi^8}{9\gamma^2 K^8}\sum_{n\neq0}n^6~\mathrm{csch}^2(n\pi).
\end{align*}
Hence,  it is immediate that $I=-\frac{d}{dc}\left\|\widehat{\phi_c}\right\|^2_{\ell^2}<0$ therefore, by Theorem~5.1 in \cite{PN} the cnoidal waves $\phi_{c}$ are also stable in $H^1_{per}\left(\left[0,L\right]\right)$.\eproof

\section{Conclusion}
In this paper,  we established the \textit{(orbital) stability} of traveling wave solutions in the case of solitary waves and periodic waves for a KdV equation which includes a  fifth order dispersive term.   We demonstrate that the sufficient conditions for stability in the current literature are satisfied for the set of traveling waves considered herein.  When the solution is given in terms of a differentiable family, the inner product $I=(\psi,\varphi)_2$ can be computed with the help of Parseval's theorem.  The inner product represents a functional constructed from conserved quantities and is used in the proof of stability theorems.  In the case where the solution does not present itself as a differentiable family, the method of Gegenbauer polynomials was used to determine the sign of the inner product $I$.  The stability is determined by applying existing results in the current literature which exploit the positivity properties of the Fourier transform of the solutions.  For periodic solutions which are given as powers of a Jacobian elliptic function we use the recurrence formula for the coefficients of the Fourier series found in \cite{AK}.  To our knowledge the stability for the traveling waves considered here have not been previously established.


\section*{References}

\begin{thebibliography}{9}
%
\bibitem{Steg} Abramowitz, M., and Stegun, I.A., (1964). Handbook of mathematical functions: with formulas, graphs, and mathematical tables (Vol. 55). Courier Corporation.
%
\bibitem{Albert} Albert, J. P. (1992). Positivity Properties and Stability of Solitary-Wave Solutions of Model Equations For Long Waves. Communications in partial differential equations, 17(1-2),  1-22.
 %
\bibitem{Albert2} Albert, J. P., and Bona, J. L. (1991). Total positivity and the stability of internal waves in stratified fluids of finite depth. IMA Journal of Applied Mathematics, 46(1-2), 1-19.
%
\bibitem{Albert3} Albert, J. P., Bona, J. L., and  Henry, D. B. (1987). Sufficient conditions for stability of solitary-wave solutions of model equations for long waves. Physica D: Nonlinear Phenomena, 24(1-3), 343-366.
%
\bibitem{PN3} Angulo Pava, J. (2003). On the instability of solitary-wave solutions for fifth-order water wave models. Electronic Journal of Differential Equations (EJDE), 2003(6), 1-18.
%
\bibitem{TB} Benjamin, T. B. (1972). The stability of solitary waves. In Proceedings of the Royal Society of London A: Mathematical, Physical and Engineering Sciences (Vol. 328, No. 1573, pp. 153-183). The Royal Society.
%
\bibitem{BBB} Bennett, D. P., Brown, R. W., Stansfield, S. E., Stroughair, J. D., and Bona, J. L. (1983). The stability of internal solitary waves. In Mathematical Proceedings of the Cambridge Philosophical Society (Vol. 94, No. 2, pp. 351-379). Cambridge University Press.
%
\bibitem{BSS} Bona, J. L., Souganidis, P. E., and  Strauss, W. A. (1987). Stability and instability of solitary waves of Korteweg-de Vries type. In Proceedings of the Royal Society of London A: Mathematical, Physical and Engineering Sciences (Vol. 411, No. 1841, pp. 395-412). The Royal Society.
%
\bibitem{JLB} Bona, J. (1975). On the stability theory of solitary waves. In Proceedings of the Royal Society of London A: Mathematical, Physical and Engineering Sciences (Vol. 344, No. 1638, pp. 363-374). The Royal Society.
%
\bibitem{BD} Bridges, T. J., and  Derks, G. (2002). Linear instability of solitary wave solutions of the Kawahara equation and its generalizations. SIAM Journal on Mathematical Analysis, 33(6), 1356-1378.
%
\bibitem{Chow} Chow, K. W., Yip, L. P., and  Grimshaw, R. (2007). Novel solitary pulses for a variable-coefficient derivative nonlinear Schr\"{o}dinger equation. Journal of the Physical Society of Japan, 76(7), 074004-074004.
%
\bibitem{Const1} Constantin, A., and  Ivanov, R. I. (2015). A Hamiltonian approach to wave-current interactions in two-layer fluids. Physics of Fluids, 27(8), 086603.

%

\bibitem{Const3} Constantin, A., and  Johnson, R. S. (2015). The dynamics of waves interacting with the Equatorial Undercurrent. Geophysical \& Astrophysical Fluid Dynamics, 109(4), 311-358.

\bibitem{Craig2} Craig, W., Guyenne, P., and  Kalisch, H. (2005). Hamiltonian long‐wave expansions for free surfaces and interfaces. Communications on Pure and Applied Mathematics, 58(12), 1587-1641.

%
\bibitem{Has} Hasimoto, H. (1970). Water waves--their dispersion and steeping. Kagaku, 40, 401-408. {\it In Japanese}
%
\bibitem{GS} Grillakis, M., Shatah, J., and  Strauss, W. (1987). Stability theory of solitary waves in the presence of symmetry, I. Journal of Functional Analysis, 74(1), 160-197.
%
\bibitem{Here} Hereman, W., Banerjee, P. P., Korpel, A., Assanto, G., Van Immerzeele, A., and  Meerpoel, A. (1986). Exact solitary wave solutions of nonlinear evolution and wave equations using a direct algebraic method. Journal of Physics A: Mathematical and General, 19(5), 607.
%
\bibitem{Karlin} Karlin, S. (1968). Total positivity (Vol. 1, p. 576). Stanford: Stanford University Press.
%
\bibitem{Kawa5} Kawahara, T. (1972). Oscillatory solitary waves in dispersive media. Journal of the physical society of Japan, 33(1), 260-264.
%
\bibitem{AK} Kiper, A. (1984). Fourier series coefficients for powers of the Jacobian elliptic functions. Mathematics of computation, 43(167), 247-259.
%
\bibitem{KdV} Korteweg, D. J., and  De Vries, G. (1895). XLI. On the change of form of long waves advancing in a rectangular canal, and on a new type of long stationary waves. The London, Edinburgh, and Dublin Philosophical Magazine and Journal of Science, 39(240), 422-443.
%
\bibitem{Man3} Mancas, S. C., and  Adams, R. (2017). Elliptic solutions and solitary waves of a higher order KdV--BBM long wave equation. Journal of Mathematical Analysis and Applications, 452(2), 1168-1181.
%
\bibitem{Mancass} Mancas, S. C. (2017).
Traveling wave solutions to Kawahara and related equations.
Differential Equations and Dynamical Systems, 19 pp.\ (DOI 10.1007/s12591-017-0367-5).
In press.
%
\bibitem{MK} McKean, H. P. (1977). Stability for the Korteweg-de Vries equation. Communications on Pure and Applied Mathematics, 30(3), 347-353.
%

\bibitem{Naga1} Nagashima, H. (1979). Experiment on Solitary Waves in the Nonlinear Transmission Line Described by the Equation $\frac{\partial  u}{\partial \tau}+u\frac{\partial u}{\partial \xi}-\frac{\partial^5 u}{\partial \xi^5}=0.$  Journal of the Physical Society of  Japan, 47(4), 1387-1388.
%
\bibitem{Natali} Natali, F. (2010). A note on the stability for Kawahara--KdV type equations. Applied Mathematics Letters, 23(5), 591-596.
%
\bibitem{PN} Pava, J. A., and  Natali, F. M. (2008). Positivity properties of the Fourier transform and the stability of periodic travelling-wave solutions. SIAM Journal on Mathematical Analysis, 40(3), 1123-1151.
%

\bibitem{PN2} Pava, J. A., Bona, J. L., and Scialom, M. (2006). Stability of cnoidal waves. Advances in Differential Equations, 11(12), 1321-1374.
%
\bibitem{Pom} Pomeau, Y., Ramani, A., and  Grammaticos, B. (1988). Structural stability of the Korteweg-de Vries solitons under a singular perturbation. Physica D: Nonlinear Phenomena, 31(1), 127-134.
%
\bibitem{Rey}  Reyes, M. A., Guti\`{e}rrez-Ruiz, D., Mancas, S. C., and Rosu, H. C. (2016). Nongauge bright soliton of the nonlinear Schr\"{o}dinger (NLS) equation and a family of generalized NLS equations. Modern Physics Letters A, 31(03), 1650020.
%
%
\bibitem{MW} Weinstein, M. I. (1986). Lyapunov stability of ground states of nonlinear dispersive evolution equations. Communications on Pure and Applied Mathematics, 39(1), 51-67.
%
\bibitem{MW2} Weinstein, M. I. (1987). Existence and dynamic stability of solitary wave solutions of equations arising in long wave propagation. Communications in Partial Differential Equations, 12(10), 1133-1173.
%
\bibitem{Yama7} Yamamoto, Y., and  Takizawa, \'{E}. I. (1981). On a solution on non-linear time-evolution equation of fifth order. Journal of the Physical Society of Japan, 50(5), 1421-1422.	

%

\end{thebibliography}
\end{document}